# Photonic hook generated by twin-ellipse microcylinder

Xingliang Shen,[1] Guoqiang Gu,[1, a)] Liyang Shao,[1, b)] Zeng Peng,[1] Jie Hu,[1] Sankhyabrata Bandyopadhyay,[1] Yuhui Liu,[1] Jiahao Jiang,[1] and Ming Chen[2]

[1] *Department of Electrical and Electronic Engineering, Southern University of Science and Technology, Shenzhen 518055, China*

[2] *Center for Information Photonics and Energy Materials, Shenzhen Institutes of Advanced Technology, Chinese Academy of Sciences, Shenzhen 518055, China*

## ABSTRACT

Recently, two special photonic jets, photonic hooks and twin photonic jets, have been proposed to deal with complex conditions in nanoscale manipulation. Photonic hooks are generated by a single light plane wave and an asymmetric microparticle, while the twin photonic jets are produced by two incident light beams. In this letter, we presented and demonstrate a method to combine photonic hooks and twin photonic jets. A single light plane wave and a symmetric microparticle, twin-ellipse microcylinder, are used in this research. The curvature degree, length and maximum $E^2$ filed enhancement of twin photonic hooks are varied significantly, with the change of refractive indices and shape of twin-ellipse microcylinder. And a liquid-immersed core-shell is built to achieve a flexible tunability.

___________________________

a) Electronic mail: gugq@sustech.edu.cn

b) Electronic mail: shaoly@sustech.edu.cn

# 1. Introduction

The electromagnetic radiation propagates along a straight line in free space is a widely accepted common sense. This truth was first challenged by Berry and Balazs [1]. They got a special solution from the linear Schrödinger equation. Such solution describes wavepackets, of which the envelope expressed by Airy functions, propagating non-spreading, self-healing and self-acceleration around a parabolic trajectory. Due to the infinite energy of Airy wave, the special solution cannot exist and have not been gotten much attentions in the past. Until 2007, Siviloglous *et al*. introduced the Airy function as the solution in optical filed [2], [3]. The optical paraxial wave function has similar form of the linear Schrödinger equation, so that Siviloglous made a reasonably good approximation and got an Airy beam with finite energy. Due to the approximation, finite energy Airy beam has a little diffraction but still has the property of self-bending transmission [2]. The self-bending beams have special potential in optical tweezers [4], super-resolution imaging [5] and light-induced curved plasma channels [6], [7]. In 2018, Yue *et al*. and Ang *et al*. proposed a new type of curved light beam, a special photonic jet which was named as 'photonic hook' [8], [9]. This curved beam, with a laser and a light modulator, is easier to generate than an Airy beam [3]. The shape and field distribution of photonic hooks are depended on the geometry and refractive index of microparticle [10]-[13]. An asymmetric microparticle of which combining a wedge prism and a cuboid was used to generate the photonic hook. The influence of the particle sizes, refractive indices and the prism angles were investigated as well. The photonic hooks were experimentally observed in terahertz range [14], and surface plasmons and acoustic waves were used to produce photonic hooks later [15], [16]. These curved beams have special use in nanoscale manipulation as optical tweezes [9]. Our previous works have shown the potential of photonic jet in single nanoparticle detection and nanoscale manipulation [17].

As for the nanoscale manipulation, single photonic jets are limited in capturing one microparticle a time, facing a trap with multiple particles. In 2017, Poteet *et al.* used two beams of light to illuminate a microscale sphere/cylinder and generate twin photonic jets for the manipulation of multiple particles [18]. The experiment and verification are completed by Chen *et al.* in 2018 [19]. Compared to the single photonic jet, not only the twin jets improve the efficiency of manipulation, the difference between two photonic jets also make them have particular use in manipulation. These twin photonic jets can significantly improve the efficiency in lithographic application.

In general, a symmetric dielectric microparticle, like sphere and ellipsoid, only produces a straight photonic jet for each incident light beam, while a bending photonic hook need an asymmetric microparticle of which illuminating by a single light beam. Hence it is hardly to generate twin photonic hooks. In this letter, we reported a method to combine photonic hook and twin photonic jets with a single incident light beam and a symmetric microparticle (twin-ellipse microcylinder). Such twin-ellipse microcylinder is created by overlapping two identical ellipse cylinders at their semi-minor axis, while the semi-major axis remains the same. Under the irradiation of plane wave, this microparticle can obviously generate double photonic jets and bending them to form photonic hooks. The curvature degrees, lengths of the photonic hooks and the inflection points can be modulated by varying the refractive indices of the twin-ellipse microcylinder and the ratios of the semi-major radius to semi-minor radius. Comparison with the traditional photonic hooks, these hooks not only have two totally same and symmetric branches, the microparticle also is symmetric, while the photonic hook from previous researches were created by an asymmetric model (combined a cuboid and a prism) [8], [9]. The key difference relative to the traditional twin photonic jets is that the two branches of these photonic hooks have identical $E^2$ filed enhancement distribution which is only illuminated by a single illuminated plane wave. The generation mechanism of twin photonic hooks from twin-ellipse microcylinder will be analyzed visually based on the energy flow distribution and the corresponding Poynting vector fields [18], [20]. When filling different liquids into a twin-ellipse microcylinder shell, the twin photonic hooks, to a certain degree, show good tunability.

## 2. Results and Discussion

This two dimensional (2D) numerical model is built in COMSOL Multiphysics commercial software package. The computational techniques are based on finite element method (FEM). Figures 1(a) and 1(b) are the three-dimensional (3D) twin-ellipse microcylinder and sectional view, respectively. The refractive index of the twin-ellipse microcylinder is $n$ surrounded by the air medium with refractive index of $n_s=1$. Ununiform meshes dependent on the refractive indices are used in the computational domain. Figure 1(c) describes the process of generating twin photonic hooks.

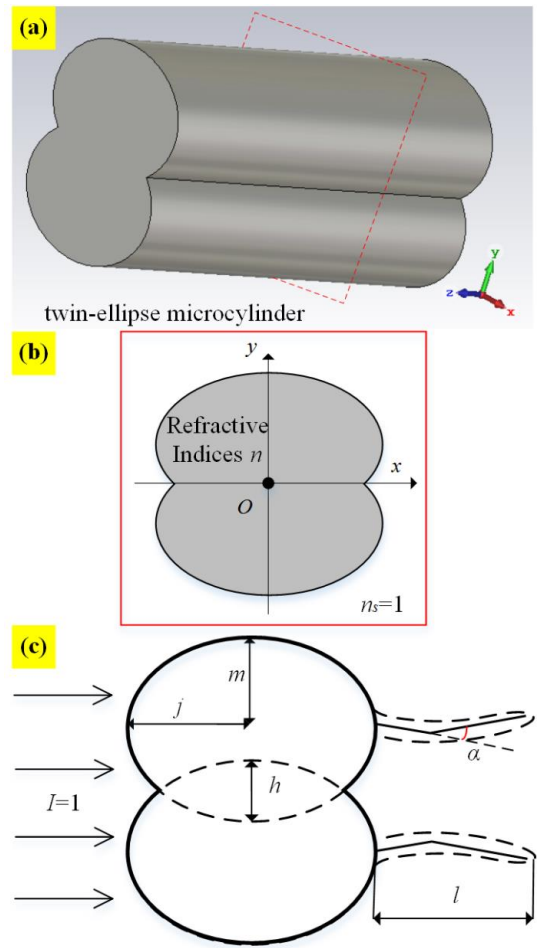

Fig. 1. (a) 3D model and (b) sectional view of twin-ellipse microcylinder. $n$ is the refractive index of the microparticle, $n_s$ is the refractive index of the surrounding medium. (c) Schematic of plane-wave-illuminated twin-ellipse microcylinder. $j$ is the semi-major radius, $m$ is the semi-minor radius, $h$ is the maximum length along $y$-axis of overlap area, $l$ is the length from output end of microparticle to $1/e$ maximum $E^2$ filed enhancement and $α$ is the inflective degree. The incident light is propagating from $-x$ to $+x$.

A plane-wave light propagates through -x to +x, with intensity $I = 1$. The semi-minor radius of a single ellipse is $m$ and the semi-major radius is $j$. The maximum length of the twin-ellipse microcylinder's overlap area, along the semi-minor axis, is $h$. $\alpha$ is the respective the bending degree of the produced photonic hooks. $l$ is the length from output end of microparticle to $1/e$ maximum $E^2$ filed enhancement. In order to avoid the influence of back reflections, the end side and the both sides of computational domain are set as perfectly marched layer absorbing boundary condition.

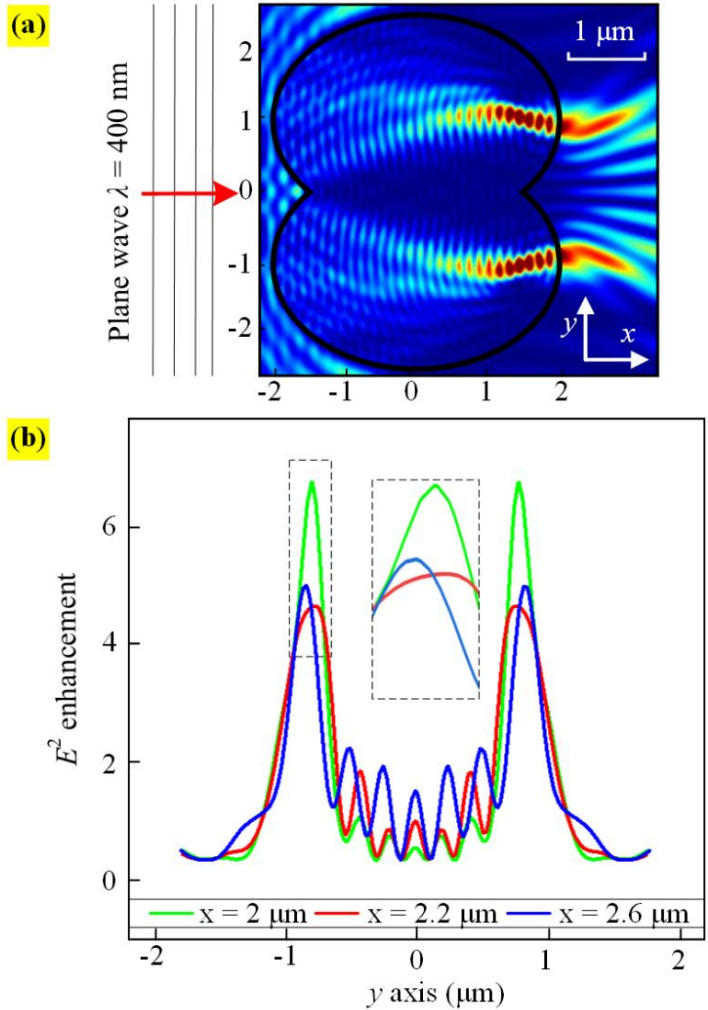

Fig. 2. (a) $E^2$ filed enhancement distribution for $j = 2$ μm, $m = 1.5$ μm, $h = 2$ μm and $n = 1.46$. (b) $E^2$ enhancement along y axis at $x = 2$ μm, $x = 2.2$ μm, $x = 2.6$ μm.

Figure 2(a) describes the $E^2$ filed enhancement distribution when a plane-wave light transmits through the twin-ellipse microcylinder with the wavelength of $\lambda = 400$ nm. The geometrical sizes and the dielectric property of the twin-ellipse microcylinder are $j = 2$ μm, $m = 1.5$ μm, $h = 1$ μm, and $n = 1.46$. It can be easily observed that, when the plane wave light illuminates the twin-ellipse microcylinder, it is gathered as jets inside of each ellipse at bottom. Then, the jets transmit through the interface, between dielectric particle and surrounding medium, at $x = 2$ μm. The light beams reach an inflection point at $x = 2.23$ μm and are bent into photonic hooks with inflective angle of $α = 140°$. The peak $E^2$ filed enhancement, outside of the twin-ellipse microcylinder, reaches 6.58 and the length $l ≈ 1$ μm. Figure 2(b) shows the $E^2$ filed enhancement distribution along the y-axis at $x = 2$ μm (green), $x = 2.2$ μm (red), and $x = 2.6$ μm (blue). The y-coordinate of the maximum $E^2$ filed enhancement drifts from $y = -0.91$ μm to $y = -0.98$ μm, then back to $y = 0.86$ μm, which can be observed explicitly in the inset of Fig.2(b). It furthermore proves the curve of the photonic hooks.

Our previous research shows the photonic jets' characters depending on the relative refractive indices of microparticle and surrounding medium, the sizes and shapes of the microparticles [17], [21]. For twin-ellipse microcylinders, the refractive indices and the ratio of major to minor radius ($j/m$) also have influences to the $E^2$ filed enhancement, the length $l$ and the inflective degree $α$. To make a visualization research, we build a series of models by varying $j/m$ at different refractive indices. Figure 3(a) shows the variations of inflective degree $α$ at three different refractive indices of $n = 1.46$ (red), $n = 1.51$ (green) and $n = 1.56$ (blue) with $m = 1.5$ μm, $h = 2$ μm and $λ = 400$ nm. In Fig.3(a), it is clear to see that the $j/m$ has a significant influence to $α$. When $j/m < 1$, the curvature degree $α$ is on a rising trend. For $n = 1.46$, the photonic hooks first appear at $j/m = 0.67$ with $α = 6.7°$, then increase to around 40°. If the $j/m$ larger than 1, $α$ shows different variation trend with the increment of $n$. For $n = 1.51$ and $n = 1.56$, $α$ declines smoothly, while the $α$ increases first and then decreases for the case of $n = 1.46$. The maximum bending angle for three different $n$ are similar. The value is about 43° around $j/m \sim 1.02$. Based on the above research, a larger $n$ brings a smaller $α$.

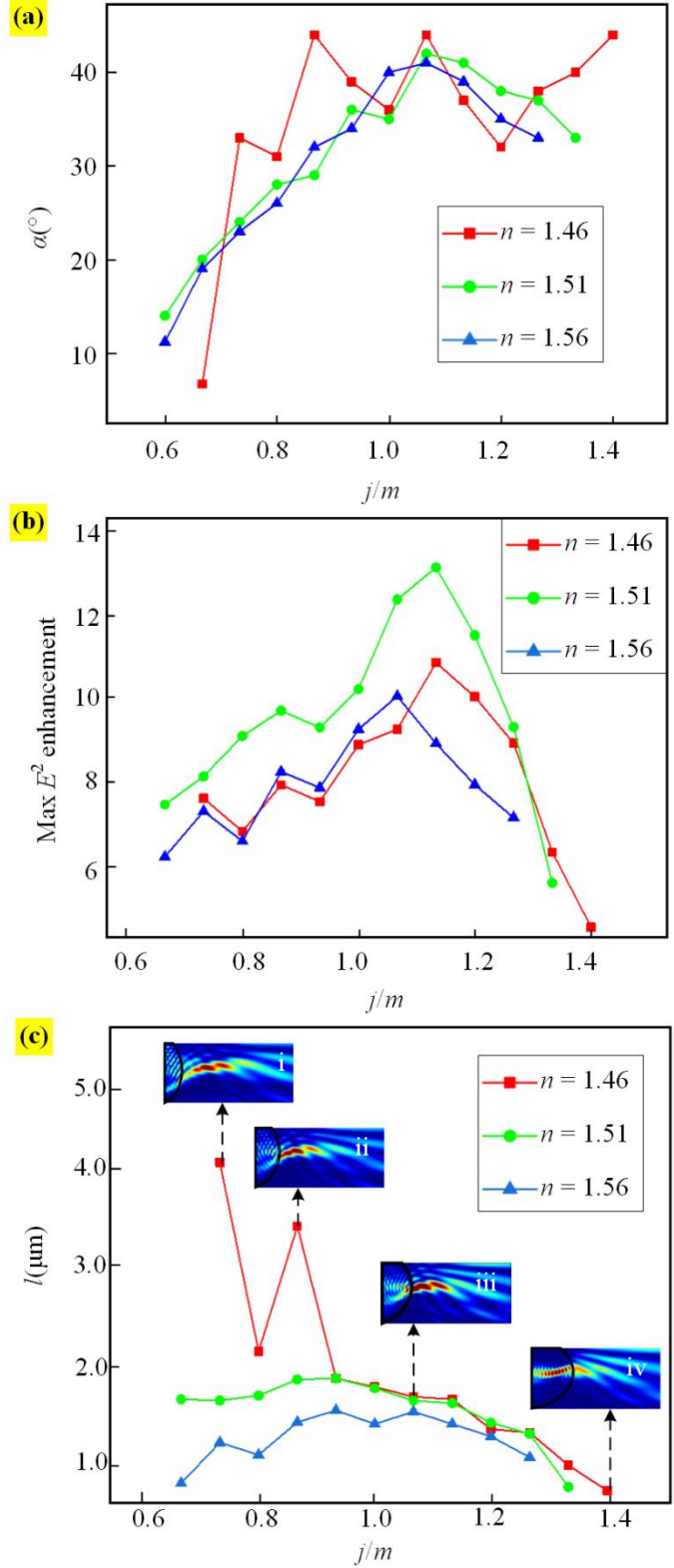

Fig. 3. (a) Curvature degree $\alpha$ (b) Maximum $E^2$ field enhancement and (c) length of photonic hooks $l$ as a function of $j/m$, where $m = 1.5$ μm, $h = 2$ μm and $\lambda = 400$ nm.

For the maximum $E^2$ filed enhancement of photonic hooks, as shown in the Fig.3(b), the variation tendencies relative to $j/m$ for the case of three different $n$ are the same. All of them have a fall after rise. The peak value is 10.9 for $n = 1.46$, 13.1 for $n = 1.51$ and 10 for $n = 1.56$, around $j/m \sim 1.1$. The maximum $E^2$ filed enhancement falls sharply after reaching peak. It is interesting to note that the absolute maximum $E^2$ filed enhancement is rising with the increasing of $j/m$. If $j/m > 1$, the points representing the absolute maximum $E^2$ field enhancement will retract back into the twin-ellipse microcylinder.

The length $l$ of the photonic hooks is also under the influence of $n$ and $j/m$ too. Figure 3(c) describes the relationship between $n$, $j/m$ and $l$. The larger $n$ brings smaller $l$. For $j/m$, $l$ declines with the rise of $j/m$, but the working principle is different when $j/m < 1$ and $j/m > 1$. The insets of Fig.3(c)-(i) and Fig.3(c)-(ii), show the electrical field distributions of the photonic hooks at $j/m = 0.73$ and $j/m = 0.86$. In this case, only the length of photonic hooks decreases influence the $l$. When $j/m > 1$, as shown in Fig.3(c)-(iii) and Fig.3(c)-(iv) for $j/m = 1.07$ and $j/m = 1.4$, the photonic hooks shorten to some extent. More importantly, the photonic hooks begin to retract into the twin-ellipse microcylinder, which shortens the $l$ more obviously.

As to the traditional photonic jets, the ratios of the refractive indices between the dielectric particle to the surrounding medium ($n/n_s$) should keep in the range of 1:1 to 2:1 [13], [22]. In the case of photonic hooks formed from the twin-ellipse microcylinder, the generation of the photonic hooks will be mainly affected by the $n/n_s$, and slightly influenced by $j/m$. Figure 4 shows the influence originating from the $j/m$ at the case of $n/n_s = 1.46$ and $h = 2$ μm. If $j/m$ is too small, as shown in Fig.4(a)-(i), the twin-ellipse microcylinder is more likely to generate photonic jets rather than photonic hooks. Figure 4(a)-(ii) describes the photonic hooks in the case of a large $j/m$, the photonic hooks will totally retract back into the twin-ellipse microcylinder. There is no hook existing in the outside of the dielectric particle. Too small $n/n_s$ brings the problem that the transmit plane wave cannot be focused into photonic jets or photonic hooks, as can be seen the case of $m$ = 1.5 μm, $j$ = 2 μm, $h$ = 2 μm in Fig.4(b)-(i). For the case of $n/n_s$=1.6 in Fig.4(b)-(ii), the hooks cannot transmit through the twin-ellipse microcylinder due to the too large focusing ability.

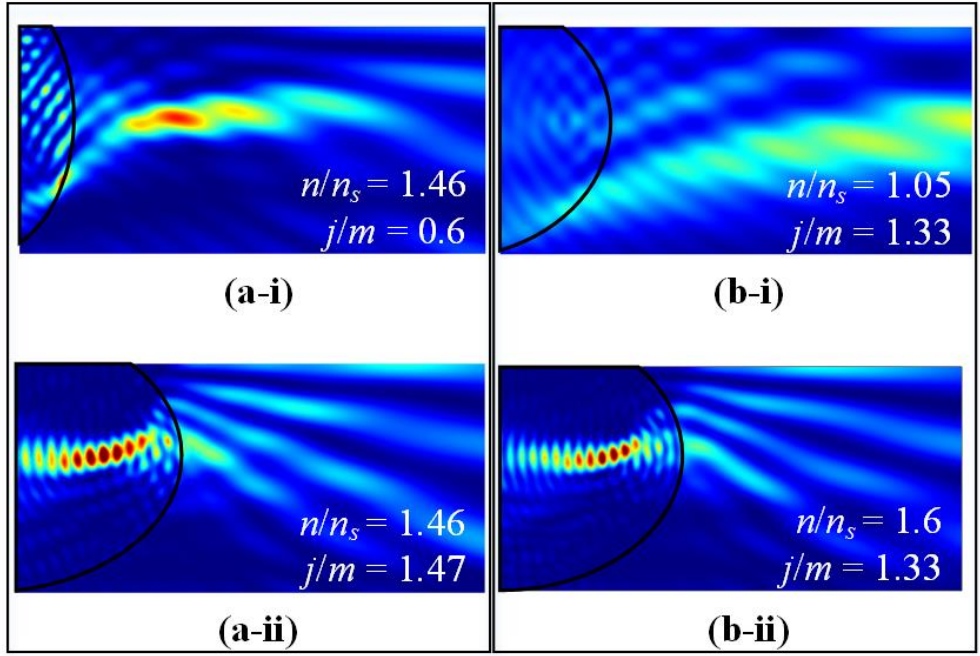

Fig. 4. $E^2$ field enhancement distribution when (a)-(1) $n/n_s$ = 1.46, $j/m$ = 0.6, (a)-(2) $n/n_s$ = 1.46, $j/m$ = 1.47, (b)-(1) $n/n_s$ = 1.05, $j/m$ = 1.33 and (b)-(2) $n/n_s$ = 1.6, $j/m$ = 1.33.

To make an analysis of the generation mechanism of the twin photonic hooks from the twin-ellipse microcylinder, the $E^2$ filed enhancement distribution and the corresponding Poynting vector fields of the microcylinder, the single-ellipse microcylinder and the twin-ellipse microcylinder are illustrated in the Fig.5. Figure 5(a) shows the simplest photonic jet formed from a microcylinder with radius of 1.5 μm, refractive index of $n$ = 1.46 and surrounding medium of air ($n_s$ = 1). It can be seen that, when the incident plane-wave transmit into the microcylinder, it will divide into two branches inside the particle and converge into a photonic jet at the rear side of the microcylinder. In the case of single-ellipse microcylinders [Fig.5(b)], the deflected beam also has two branches before the focused photonic jet appears.

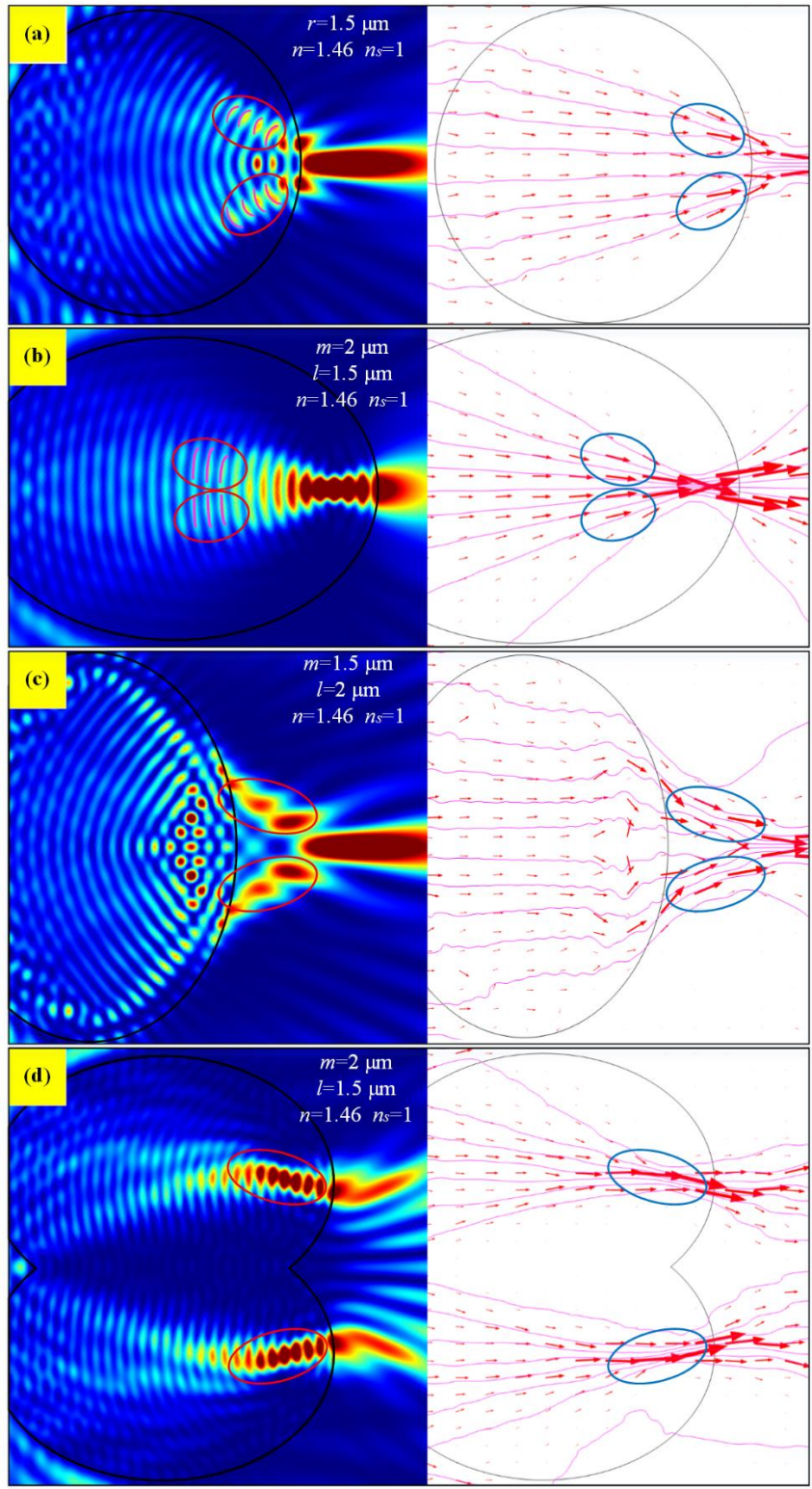

Fig. 5. $E^2$ field enhancement distribution, Poynting vector distributions (pink line) and energy flow streamlines (red arrows) of (a) microcylinder, (b) single-ellipse microcylinder, (c) vertical single-ellipse microcylinder and (d) twin-ellipse microcylinder.

For a vertical single-ellipse microcylinder, as shown in Fig.5(c), these two branches are bending in the outside of the microparticle. After the two bending beams merging into a jet, the inflection of transmission beams disappears. These three figures illustrate that the dielectric particle will converge the incident light as two bending branches propagating to the bottom of microparticle. Because of the symmetric of microcylinder and single-ellipse microcylinder, the two branches are totally the same. The curvatures will cancel out while two branches make up as a straight photonic jet. But for the twin-ellipse microcylinders, the overlap of two ellipse prevent the merging of two branches. As shown in Fig.5(b), each ellipse only inducing one curved focusing beam. This beam can transmit through the interface of microparticle's bottom and surrounded medium without interference with other beams. It is probably the reason of the twin photonic hooks formed from the twin-ellipse microcylinder.

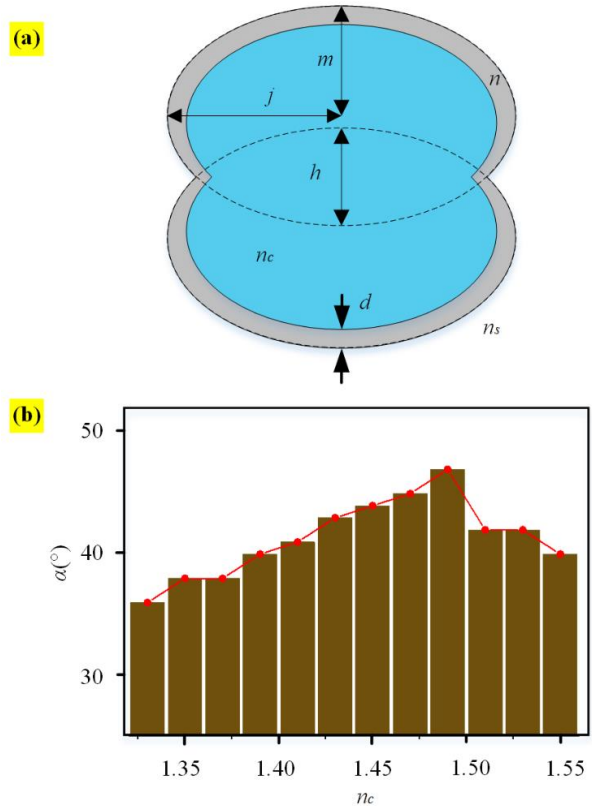

Fig. 6. (a) Schematic of liquid-immersed core-shell twin-ellipse microcylinder with semi-minor radius $m$, semi-major radius $j$, the maximum length along $y$-axis of overlap area $h$, thickness of shell $d$ and the refractive indices of core, shell, and surrounding medium $n_c$, $n$, $n_s$. (b) The curvature degree $\alpha$ with $n_c$ changing from 1.33 to 1.55.

Referring to the researches in Ref.10 and 23, we propose a scheme to make the photonic hooks from twin-ellipse microcylinder turnability. According to the preview works, $n/n_s$ can significantly affect the curvature degree, the maximum $E^2$ filed enhancement and the length of the photonic hooks. We built a liquid-immersed core-shell twin-ellipse microcylinder in theory [24]. By varying the refractive indices of the liquid filled in the core, the photonic hooks can be modified easily. Figure 6(a) describes the sectional view of this hollow-core twin-ellipse microcylinder. The $m$ and $j$ are the semi-minor radius and semi-major radius of shell, $d$ is the thickness of the shell, $h$ is maximum length along $y$-axis of the overlap area. The refractive indices of core and shell is $n_c$ and $n$, and the surrounding medium is air with $n_s = 1$. Figure 6(b) shows the bending angles of photonic hooks with variational $n_c$ when $m = 1.5$ μm, $j = 1.8$ μm, $h = 2$ μm, $d = 0.15$ μm and $n = 1.46$ with incident light wavelength of $\lambda = 400$ nm. $\alpha$ is increasing from $n_c = 1.33$ to $n_c = 1.49$, and falls until $n_c = 1.55$. The maximum $\alpha$ is 46°.

## 3. Conclusion

In conclusion, this paper conducted a combination research between the twin photonic jets and the photonic hooks. A symmetric microparticle, twin-ellipse microcylinder, and single beam are used in this study. It is different from only generating twin photonic jets or single photonic hooks. By changing the refractive indices contrasts and the ratios between semi-major and semi-minor radius, the curved degree, intensity and length of the photonic hooks can be significantly affected. With the analysis of Poynting vector fields, the generation mechanism of the twin photonic hooks is visible. A modulation method, creating a liquid-immersed core-shell twin-ellipse microcylinder, is also proposed to flexible adjust the curvature degrees. These twin photonic hooks have great potential in complex nanoscale manipulation with multiple particles, and its identical branches also have particular use in super resolution imaging.

## Author Information

### Author Contributions

Xingliang Shen and Guoqiang Gu contributed equally to this work.

### Notes

The authors declare no competing financial interest.

## Acknowledgements

This work was supported by the Guangdong Basic and Applied Basic Research Foundation (2019A1515011242), Shenzhen Postdoctoral Research Grant Program (K19237504), and the startup fund from Southern University of Science and Technology and Shenzhen government.